\documentclass[aps,pre,twocolumn,superscriptaddress,showpacs]{revtex4}
\usepackage{graphicx}
\RequirePackage{times}

\begin{document}

\title{Computation of the temperature dependence of the heat capacity of complex molecular
systems using random color noise}

\author{Sahin Buyukdagli}

\affiliation{Department of Physics and Centre for Nonlinear Studies,
Hong Kong Baptist University, Hong Kong, China}

\author{Alexander V. Savin}
\affiliation{Semenov Institute of Chemical Physics, Russian Academy
of Sciences, Moscow 119991, Russia}

\author{Bambi Hu}
\affiliation{Department of Physics and Centre for Nonlinear Studies,
Hong Kong Baptist University, Hong Kong, China}
\affiliation{Department of Physics, University of Houston, Houston,
Texas 77204-5005}

\begin{abstract}
We propose a new method for computing the temperature dependence of
the heat capacity in complex molecular systems. The proposed scheme
is based on the use of the Langevin equation with low frequency
color noise. We obtain the temperature dependence of the correlation
time of random noises, which enables to model the partial
thermalization of high-frequency vibrations, which is a pure quantum
effect. By applying the method to carbon nanotubes, we show that the
consideration of the color noise in the Langevin equation allows to
reproduce the temperature evolution of the specific heat with good
accuracy.
\end{abstract}


\maketitle

\begin{widetext}
\tableofcontents

~\\
\end{widetext}

\section{Introduction}
The pronounced temperature dependence of the specific heat $c(T)$ in
molecular systems is a pure quantum effect. It is well known that in
the absence of any critical point, the increase of the temperature
is accompanied by a smooth rise of the specific heat $c(T)$, while
by decreasing the temperature, $c(T)$ tends to zero. The first
explanation for this phenomenon was given by Einstein a century ago
\cite{en1}. This quantum effect results from the fact that at low
temperatures, high frequency vibrations are partly frozen while low
frequency vibrations are fully excited. It is of course impossible
to explain the effect of partial thermalization within the framework
of classical physics. In fact, the use of the usual Langevin
equation with white noise leads to a uniform thermalization of all
modes and the specific heat is practically insensitive to the
temperature (in classical physics, any temperature dependence of the
thermal capacity results from non-linearity effects). On the other
hand, it is possible to mimic the partial thermalization effect in
the framework of the Langevin description if one considers, instead
of a white noise, a low frequency color noise with a temperature
dependent frequency spectrum. To this end, it is enough to introduce
a random  noise with a finite correlation time $t_c>0$ during which
the noise "remembers" \ its last realization. In this study, we
obtain the temperature dependence of this correlation time, which
allows a correct modeling of the partial thermalization of
vibrations in the system.

The paper is organized as follows. We consider in
Sec. I a harmonic oscillator and investigate its dynamics described
by the Langevin equation with color noise. We obtain the temperature
dependence of the correlation time of random noises, which enables
us to efficiently model the partial thermalization of high-frequency
vibrations. We next evaluate in Sec. II the effect of
non-linearities on the accuracy of our computational method and
propose a general scheme to compute the heat capacity for many-body
systems using a color noise. The final section is devoted to the
application of the proposed scheme to a Hamiltonian carbon nanotube
model.  It is shown in this section that the temperature evolution
of the specific heat computed with the Langevin equation with color
noise agrees well with the result obtained from a direct quantum
mechanical calculation.

\section{The Langevin equation}

The thermalization of the mode of frequency $\Omega$ is described by
the Langevin equation
\begin{equation}
\ddot{u}+\Omega^2 u+\Gamma\dot{u} = \xi(t)/\mu~ ,
\label{f1}
\end{equation}
where $u$ is the coordinate of the vibration; damping
$\Gamma=1/t_r$, and $t_r$ is the relaxation time; $\mu$ is the
reduced mass of the mode; $\xi(t)$ is a normally distributed
random force which describes the interaction of the mode with the
thermal bath of temperature $T$ and the autocorrelation function
$$
\langle \xi(t)\xi(t') \rangle = 2\mu \Gamma k_B T\varphi(t-t')
$$
(the dimensionless function $\varphi (t)$ is normalized as
$\int_{-\infty}^{+\infty}\varphi (t)dt=1$), where $k_B$ is the Boltzmann constant.

At thermal equilibrium the averaged energy of thermal vibrations is
defined by the relation
\begin{eqnarray}
E=\lim_{\tau\rightarrow\infty}\frac{1}{\tau}
\int_{0}^{\tau}\frac{\mu}{2}(\dot{u}^2+\Omega^2u^2)dt \nonumber\\
=\int_{0}^{+\infty}\mu
(\omega^2+\Omega^2)\left| H(\omega)\right|^2 F(\omega)d\omega~,\label{f2}
\end{eqnarray}
where $H(\omega)=\left[
\mu(\Omega^2-\omega^2+i\omega\Gamma)\right]^{-1}$ is the transmission
function and $F(\omega)$ is the Fourier transform of the
autocorrelation function of the random force,
\begin{eqnarray}
F(\omega)=\frac{1}{2\pi}\int_{-\infty}^{+\infty}\langle \xi(t)\xi(0)\rangle
\exp \{ -i\omega t\} dt\nonumber\\
=\frac{\mu \Gamma k_BT}{\pi}\int_{-\infty}^{+\infty}\varphi(t)\exp
\{-i\omega t \} dt~. \label{f3}
\end{eqnarray}
Thus, $E=K+P$ where the averaged kinetic energy is defined by the relation
\begin{eqnarray}
K=\lim_{\tau \rightarrow\infty}\frac{1}{\tau}\int_{0}^{\tau}\frac{1}{2}\mu
\dot{u}^2 dt\nonumber\\
=2k_BT\Gamma\int_{0}^{+\infty}\frac{\omega^2{\cal F}(\omega)d\omega}
{(\Omega^2-\omega^2)^2+\omega^2\Gamma^2}
\label{f4}
\end{eqnarray}
and the averaged potential energy is given by
\begin{eqnarray}
P=\lim_{\tau \rightarrow\infty}\frac{1}{\tau}\int_{0}^{\tau}\frac{1}{2}
\mu\Omega^2 u^2 dt\nonumber\\
=2k_BT\Gamma\int_{0}^{+\infty}\frac{\Omega^2{\cal F}(\omega)d\omega}
{(\Omega^2-\omega^2)^2+\omega^2\Gamma^2}~,
\label{f5}
\end{eqnarray}
where ${\cal F}(\omega )$ is the Fourier transform of the dimensionless
autocorrelation function $\varphi (t)$:
$$
{\cal F}(\omega )=
\frac{1}{2\pi}\int_{-\infty}^{+\infty}\varphi (t)\exp \{-i\omega t\} dt~.
$$

For a delta-correlated random force (the case of the white noise),
$\varphi (t)=\delta(t)$ and ${\cal F}(\omega )=1/2\pi$.
The integrals (\ref{f4}) and (\ref{f5}) can be easily calculated by
a contour integration. The energy $E=K=k_BT/2$ at $\Omega=0$ and
$E=K+P=k_BT$, $K=P=k_BT/2$ for frequency $\Omega >0$.

For an exponentially-correlated random force (the case of low
frequency color noise), $\varphi(t)=\frac12 \lambda\exp{-|\lambda
t|}$ and ${\cal F}(\omega)=\lambda^2/2\pi (\omega^2+\lambda^2)$,
where $\lambda =1/t_c$ and $~t_c$ is the correlation time of the
random force. In this case, the integrals (\ref{f4}) and (\ref{f5})
can also be calculated by a contour integration. The averaged
kinetic energy $K=k_BTf_K(\Omega,\Gamma,\lambda)/2$ and the averaged
potential energy $P=k_BTf_P(\Omega,\Gamma,\lambda)/2$ with $f_K$ and
$f_P$ defined by

\begin{eqnarray}
f_K(\Omega,\Gamma,\lambda) &=&
\lambda^2/(\lambda^2+\lambda\Gamma+\Omega^2)~,\label{f6}\\
f_P(\Omega,\Gamma,\lambda) &=&
\frac{\lambda^2(\Omega^2+\lambda^2-\Gamma^2)+\lambda\Gamma\Omega^2}
{(\Omega^2+\lambda^2)^2-\Gamma^2\lambda^2}~.\label{f7}
\end{eqnarray}

In the case of an harmonic oscillator, the Langevin equation with white noise
describes thermal vibrations of harmonic modes in the classical approximation,
where the mean energy obeys $E=k_BT$. In the case of a quantum
harmonic oscillator $H=\hbar\Omega(B^+B+\frac12)$, where $\hbar$ is
the Planck constant, $B^+$ and $B$ represent creation and
annihilation operators, the mean energy of thermal vibrations is
given by

\begin{equation}
E(\Omega,T)=\frac{\hbar\Omega}{\exp(\hbar\Omega/k_BT)-1}+\frac12\hbar\Omega~.
\label{f8}
\end{equation}
The heat capacity of the oscillator is defined by
$c(\Omega,T)=dE(\Omega,T)/dT=k_BF_E(\Omega,T)$, where the Einstein
function behaves according to
$$
 F_E(\Omega,T)=\left(\frac{\hbar\Omega}{k_BT}\right)^2
 \frac{\exp(\hbar\Omega/k_BT)}{[\exp(\hbar\Omega/k_BT)-1]^2}~.
$$
For $T\rightarrow \infty$, the Einstein function
$F_E(\Omega,T)\rightarrow 1$, while in the limit $T\rightarrow 0$,
we get $F_E(\Omega,T)\rightarrow 0$. Consequently at high
temperatures, the specific heat behaves as $c(\Omega,T)\approx k_B$
and at low temperatures, one has $c(\Omega,T)\approx 0$. For this
reason, in the low temperature regime defined by
$T<T_E=\hbar\Omega/k_B$, vibrations of the quantum oscillator are
partially frozen. Consequently, a classical description of thermal
vibrations is valid exclusively in the high temperature regime
$T>T_E$, where the Einstein temperature $T_E$ is defined by
$F_E(\Omega,T_E)=e/(e-1)^2=0.9206735$ .

If we drop the energy of vacuum vibrations $\hbar\Omega/2$, the thermalization of
the quantum oscillator can be characterized by the function

\begin{eqnarray*}
G(\Omega,T)=[E(\Omega,T)-\hbar\Omega/2]/k_BT\nonumber\\
=\frac{\hbar\Omega/k_BT}{\exp(\hbar\Omega/k_BT)-1}~.
\end{eqnarray*}
In the limit $T\rightarrow 0$, the thermalization coefficient
$G(\Omega,T)\rightarrow 0$, while at $T=T_E$ the function
$G(\Omega,T_E)=1/(e-1)=0.5819767$, and in the limit
$T\rightarrow\infty$, we have $~G(\Omega,T)\rightarrow 1$. Hence for
temperatures $T>0$, vibrations with frequency
$\Omega>\Omega_E(T)=k_BT/\hbar$ will be frozen. We can thus conclude
that it is uncorrect to model thermal fluctuations of these modes
using the Langevin equation with white noise.

As we stated at the beginning of this paper, the partial
thermalization of high frequency vibrations and the total
thermalization of low frequency modes can be realized if one uses
a Langevin equation with color noise which consists of low frequency components of
the white noise. But it is necessary to consider in this case the temperature dependence
of the noise correlation functions. For an exponentially-correlated random noise,
this dependence can be deduced  from the relation

$$
G(\Omega_E(T),T)\approx [f_K(\Omega_E(T),\Gamma,\lambda)+f_P(\Omega_E(T),\Gamma,\lambda)]/2.
$$
In the  limit $\Gamma\ll \Omega_E(T)=k_BT/\hbar$, using Eq. (\ref{f6}) and (\ref{f7})
the last equation can be expressed in the simple form

\begin{equation}
\frac{\lambda^2}{\lambda^2+(k_BT/\hbar)^2}=\frac{1}{e-1}~.
\label{f9}
\end{equation}
The equation (\ref{f9}) yields the following linear temperature
dependence of the correlation coefficient :

\begin{equation}
\lambda=1/t_c=k_BT/\hbar\sqrt{e-2}~.\label{f10}
\end{equation}

It follows from Eq. (\ref{f10})  that the description of the partial
thermalization of high frequency vibrations with the use of the
Langevin equation (\ref{f1}) becomes possible if we introduce a
correlation time $t_c$ that is inversely proportional to the
temperature $T$ of the thermal bath, i.e.

\begin{equation}
t_c=\hbar\sqrt{e-2}/k_BT~.\label{f11}
\end{equation}

The time correlated color noise can be inserted in the Langevin
description if one replaces the Langevin equation (\ref{f1})
by the system of two equations,

\begin{eqnarray}
\ddot{u}&=&-\Omega^2u-\Gamma\dot{u}+\xi(t)/\mu, \label{f12}\\
\dot{\xi} &=& (\eta(t)-\xi(t))/t_c, \label{f13}
\end{eqnarray}
where $\eta(t)$ stands for the white noise generator, normalized according to

$$
\langle\eta(t)\eta(t')\rangle=2\mu\Gamma k_BT\delta(t-t'),
$$
and $t_c$ is the correlation time whose temperature dependence is determined by (\ref{f11}).
\begin{figure}[b]
\includegraphics[angle=0, width=1\linewidth]{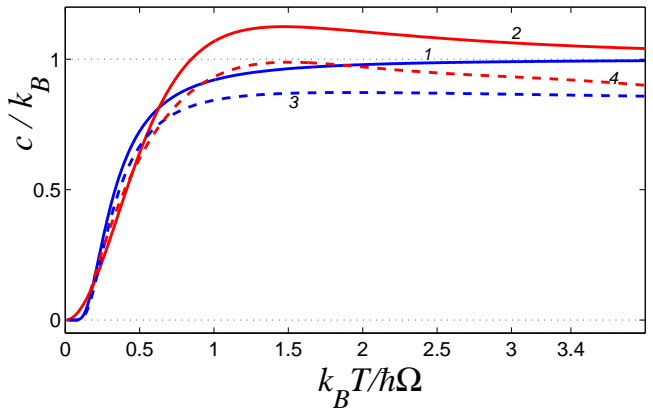}
\caption{Temperature dependence of the heat capacity of the quantum
harmonic (curve 1) and anharmonic oscillator (curve 3), classical
harmonic (curve 2) and anharmonic oscillator (curve 4) obtained from
the Langevin equation with color noise ($\Omega$ is the frequency of
the oscillator and the anharmonicity parameter is chosen as
$\beta=0.1$).} \label{fig1}
\end{figure}

Fig. \ref{fig1} compares the specific heat of a quantum harmonic oscillator
and  a classical oscillator whose dynamics is described by Langevin
equations with color noise (\ref{f12}) and (\ref{f13}).
It is clear that the introduction of the color noise doesn't allow
to reproduce the temperature dependence of the quantum oscillator exactly.
This is mainly due to the well known inadequacy of a classical description
to describe zero vibrations (ground state fluctuations). The figure
nevertheless shows that the use of the color noise yields
a qualitatively correct behaviour of the heat capacity, that is, in the limit
$T\rightarrow 0$ the specific heat
$c\rightarrow 0$, and for $T\rightarrow\infty$ one has $c\rightarrow k_B$
(with the increase of the temperature, the correlation time tends
to zero and the color noise becomes a white noise).
Most importantly, we notice that there is a good agreement
between the exact quantum result and the modified Langevin description
at low temperatures $T<T_E$ where quantum effects dominate.

We have shown that by using the Langevin equation with color noise,
one can obtain a qualitatively correct picture for
the temperature evolution of the specific heat in the case of a harmonic
oscillator. The presence of non-linearities in the system
will be tackled in the next section.

\section{The efficiency of the proposed scheme in the presence of non-linearities}

As is well-known, anharmonicities are always present in physical
systems and the validity of the harmonic approximation which
consists in representing the building blocks of a condensed system
by linear oscillators is usually restricted to very low energy
regimes. The anharmonic intermolecular forces may result either from
the non-linearity of the individual oscillators or from their
non-linear mutual interaction. The natural question arises whether
the Langevin equation with color noise can be used in the presence
of non-linearities in the system. As a first attempt to answer this
question, let us consider a non-linear oscillator whose
dimensionless Hamiltonian is given by
\begin{equation}
H=\frac12\dot{u}^2+\frac12u^2+\frac14\beta u^4~.
\label{f14}
\end{equation}
The corresponding Langevin equation with color noise can be written
in the form
\begin{eqnarray}
\ddot{u}&=&-u-\beta u^3-\Gamma\dot{u}+\xi(t), \nonumber\\
\dot{\xi} &=& (\eta(t)-\xi(t))/t_c, \label{f15}\\
&&\langle\eta(t)\eta(t')\rangle=2\Gamma T\delta(t-t'),\nonumber
\end{eqnarray}
where $T$ - the dimensionless temperature, the friction coefficient
$\Gamma=0.01$ and the correlation time $t_c=\sqrt{e-2}/T$.

The numerical integration of the set of equations of motion
(\ref{f15}) yields the mean energy $E=\langle H\rangle$ of the
anharmonic oscillator as a function of $T$. Then the specific heat
is computed from $c(T)=dE/dT$.

In order to check the accuracy of the modified Langevin equation, we
equally obtained the specific heat of this oscillator by computing numerically
the  exact eigenvalues $E_n$ of the quartic Hamiltonian (\ref{f14}). The diagonalization
of the hamiltonian matrix was performed in the basis of the harmonic oscillator.
The obtained eigenvalues were then used to find the partition function and the specific heat
from the well-known relations

\begin{eqnarray}
Z(T)&=&\sum_n e^{-E_n/T}, \nonumber\\
F(T)&=&-T\ln(Z), \label{f15.5}\\
c(T)&=&-T\frac{d^2F}{dT^2}.\nonumber
\end{eqnarray}

We compare in Fig. \ref{fig1} the result of the numerical simulation
to that obtained from the quantum statistical calculation (curve 3
and 4) at $\beta=0.1$. We notice that the non-linearity effects are
indistinguishable within the accuracy of the proposed method.

Let us note that Kleinert`s variational path integral method allows
us to obtain a fully analytical expression for the specific heat of
this quantum oscillator \cite{Kleinert}. The method aims at approximating
the quantum partition function

\begin{equation}\label{QUPART}
Z=\int\mathbf{\emph{D}x}\hspace{0.5mm}e^{-S[\mathbf{x}]},
\end{equation}
where

\begin{equation}\label{Action}
S=\int d\tau\left(\frac{m}{2}\dot{\mathbf{x}}^2+V[\mathbf{x}]\right)
\end{equation}
is the Euclidean action, with a quadratic trial action
$S_0[\mathbf{x}]$ whose variational parameters are fixed by minimizing
the right hand sight of the Jensen-Peirels inequality,

\begin{equation}\label{Jensen}
F\leq F_0 +T\left<S-S_0\right>.
\end{equation}
This first order perturbation theory allows to compute the approximative
quantum partition function of a $N$-body system as an integration
over the classical configurational space,

\begin{equation}\label{f15.6}
Z=\int\frac{d\mathbf{x_0}}{\left(2\pi/T\right)^{N/2}}e^{-W(\mathbf{x_0})/T}.
\end{equation}
where $W(\mathbf{x_0})$ is the so-called \textit{centroid potential} and $x_0$
stands for the centroid path \cite{Kleinert,Tognetti}.

The effective potential $W(x_0)$ obtained by Kleinert for the
trivial case of the quartic oscillator \cite{Kleinert} is given in the
appendix. We verified that for $\beta=0.1$, the analytical
expression of the specific heat  obtained from the partition
function (\ref{f15.6}) reproduces the numerical result (curve 3) of
Fig. \ref{fig1} with an error that is imperceptible on this
scale.

In order to examine the effect of non-linearities present in mutual
interactions between vibrational modes, let us consider now the case
of two linear oscillators with a dimensionless Hamiltonian

\begin{equation}
H=\frac12(\dot{u}_1^2+\dot{u}_2^2+\omega_1^2u_1^2+\omega_2^2u_2^2)+\frac14\beta(u_1-u_2)^4,
\label{f16}
\end{equation}
where the variable $u_i$ describes the dynamics of the  $i$-th
oscillator $(i=1,2)$, $\omega_i$ the frequency of the same mode, and
the parameter $\beta$ sets the strength of the non-linear
interaction between the two oscillators. For this two-body
hamiltonian, the system of Langevin equations with color noise takes
the form

\begin{eqnarray}
\ddot{u_1}&=&-\omega_1u_1+\beta(u_2-u_1)^3-\Gamma\dot{u_1}+\xi_1(t), \nonumber\\
\ddot{u_2}&=&-\omega_2u_2-\beta(u_2-u_1)^3-\Gamma\dot{u_2}+\xi_2(t), \label{f17}\\
\dot{\xi_1} &=& (\eta_1(t)-\xi_1(t))/t_c, \nonumber \\
\dot{\xi_2} &=& (\eta_2(t)-\xi_2(t))/t_c, \nonumber
\end{eqnarray}
where $\eta_i(t)$ stands for the random function that generates
white noise with normalization conditions
$$
\langle\eta_i(t)\eta_i(t')\rangle=2\Gamma T\delta(t-t'),~~i=1,2,~~
\langle\eta_1(t)\eta_2(t')\rangle=0.
$$
($T$ - dimensionless temperature, the  dissipation coefficient is
chosen as $\Gamma=0.01$, and the correlation time is given by
$t_c=\sqrt{e-2}/T$).

It is also possible for this two-body system to obtain the
temperature dependence of the specific heat from a purely quantum
mechanical calculation, that is, by diagonalizing the hamiltonian
matrix corresponding to Eq. (\ref{f16}) in the product basis of two
harmonic oscillators of frequencies $\omega_1$ and $\omega_2$. The
obtained energy levels were then used in Eq. (\ref{f15.5}) in order
to obtain the specific heat.

\begin{figure}[t,b]
\includegraphics[angle=0, width=1\linewidth]{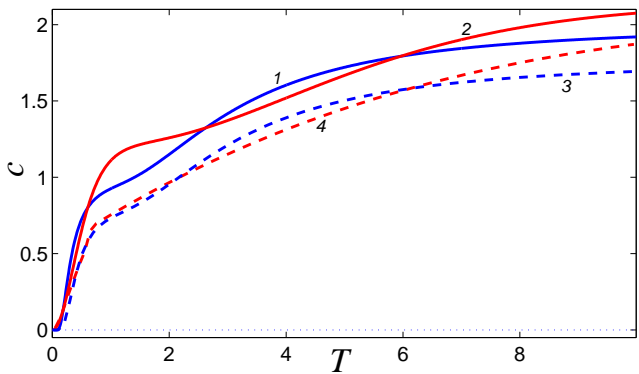}
\caption{ Temperature dependence of the heat capacity for the system of two
coupled linear oscillators (with frequencies $\omega_1=1$,
$\omega_2=10$) obtained from a direct quantum mechanical calculation
(curves 1, 3) and the Langevin equation with color noise (curves 2,
4). Solid lines correspond to decoupled oscillators ($\beta=0$) 
and dashed lines to a coupling $\beta=1$. } \label{fig2}
\end{figure}

For  the sake of distinctness, we chose $\omega_1=1$ and
$\omega_2=10$. The result of the numerical integration of the
equations of motion (\ref{f17}) illustrated in Fig. \ref{fig2}
indeed shows that the presence of non-linearities in the interaction
between individual modes improves the precision of the proposed
Langevin approach with color noise. To be more precise, for an
interaction strength of $\beta=1$, the temperature dependence of
$c(T)$ agrees better than the non-interacting case $\beta=0$ with
the result obtained from quantum statistical mechanics.

Using once more Kleinert`s variational path integral method, we
derived an analytical expression for the partition function of the
two-body Hamiltonian (\ref{f16}). The calculation is summarized in
the appendix and the analytical result agrees very well with the
numerical result in the temperature regime $T>0.3$.

These two examples clearly shows that the use of the Langevin
equation with color noise can be used as a reasonably accurate tool
to compute the specific heat of complex molecular systems that are
very difficult to study within the framework of quantum statistical
mechanics.

We will now present the scheme for evaluating the temperature
dependence of the specific heat for general molecular systems by
using a color noise. Let us define the $N$ dimensional vector ${\bf
x}=\{x_n\}_{n=1}^N$ which denotes the spatial coordinates of the
individual atoms in the system. In terms of these coordinates, the
generalized Langevin equation describing the dynamics of the system
takes the form

\begin{eqnarray}
{\bf M}\ddot{\bf x} &=& -\partial H/\partial{\bf x}-\Gamma {\bf M}\dot{\bf x}+\Xi,
\label{f18}\\
\dot{\Xi} &=& (\Theta-\Xi)/t_c,
\label{f19}
\end{eqnarray}
where ${\bf M}$ - the mass matrix of the atoms, $H$ - the
Hamiltonian of the system, the dissipation coefficient is given by
$\Gamma=1/t_r$ ($t_r$ - the relaxation time),
$\Theta=\{\eta_n\}_{n=1}^N$ is the vector of normally distributed
random forces obeying the normalization conditions

\begin{equation}
\langle \eta_n(t_1)\eta_l(t_2)\rangle= 2M_nk_BT\delta_{nl}\delta(t_1-t_2),
\label{f20}
\end{equation}
and the temperature dependence of the correlation time $t_c$ is
still determined by the relation (\ref{f11}). The choice of
characteristic times for the integration of the generalized Langevin
equations requires some care. In fact, since the formula (\ref{f11})
was obtained in the limit $\Gamma\gg k_BT/\hbar$, the numerical
value of the relaxation time should be large enough. It is thus
crucial that the fixed value of the relaxation time remains always
bigger than the correlation time of random forces in the considered
temperature regime. On the other hand, very large values of $t_r$
are not suitable since it would require exceedingly long integration
times to drive the system to thermal equilibrium. From a practical
point of view, the numerical value $t_r=1$ ps is adequate since the
result remains practically invariant under the increase of this
characteristic time.

The proposed scheme will be applied in the next section to a
Hamiltonian carbon nanotube model.

\section{Computation of the heat capacity of carbon nanotubes}

Carbon nanotubes have the peculiarity of behaving as
quasi-one-dimensional systems. This characteristic allows us to
compute the specific heat of these structures using quantum
statistical tools and this is what we will exploit in this section
in order to check the efficiency of our modified Langevin approach
on a concrete molecular system.

\begin{figure*}[t,b]
\includegraphics[angle=0, width=1\linewidth]{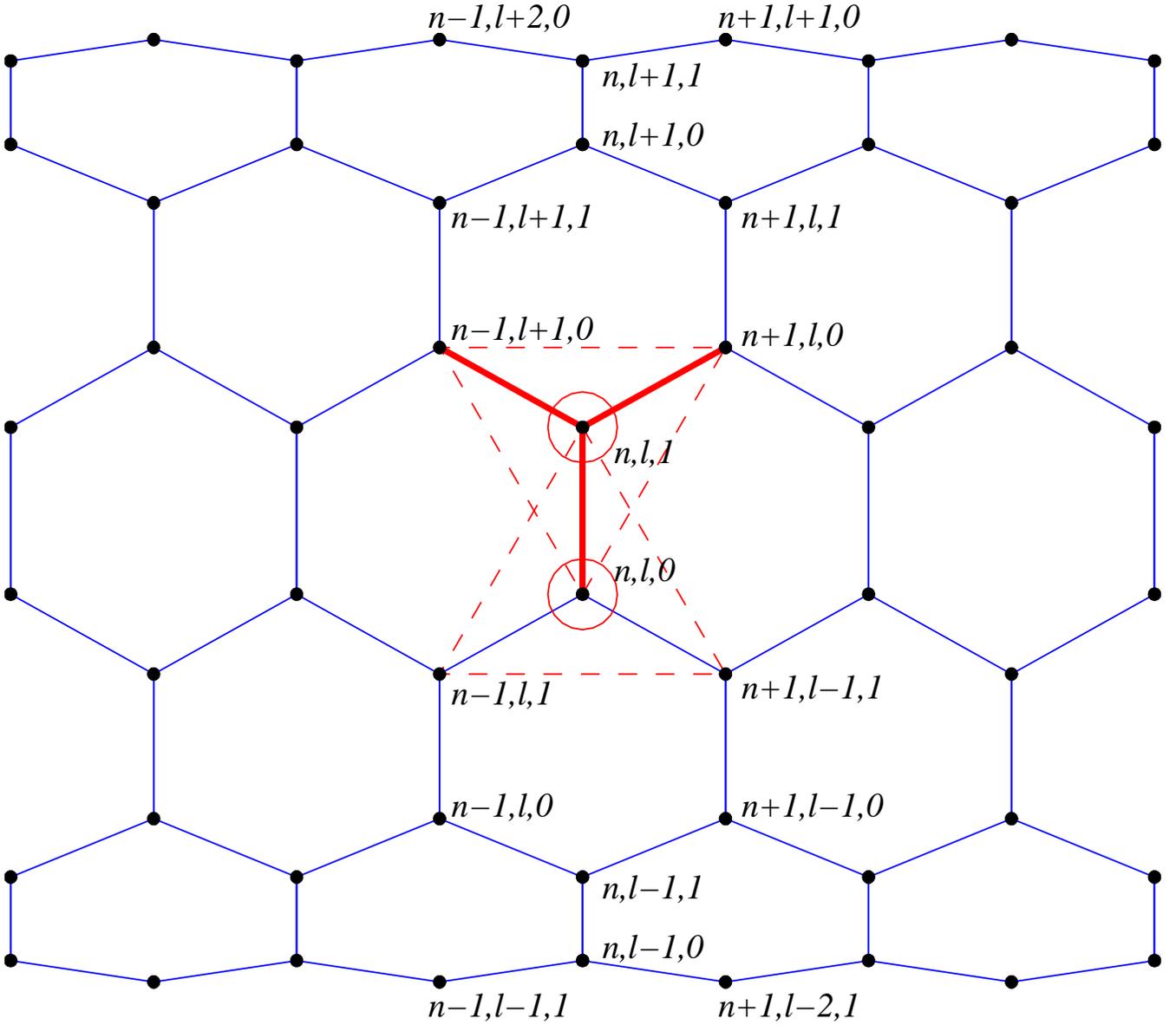}
\caption{
Schematic representation of a carbon nanotube with
chirality $(m,m)$ and numbering of atoms in the structure. Thick red
lines mark valent bond couplings, thin red arcs marks valent angle
couplings, thin dashed lines show the foundation of two pyramids
which form dihedral angles along the valent bonds in the elementary
cell $(n,l)$. For this figure, we chose $m=6$.
}
\label{fig3}
\end{figure*}

For the sake of simplicity we will limit ourselves to the case of
a nanotube with index of chirality $(m,m)$.
The structure of a carbon nanotube (CNT) with
chirality $(m,m)$ (armchair structure) is shown schematically in
Fig.~\ref{fig3}. The nanotube is characterized by its radius $R$, the
angle shift $\varphi$ and the longitudinal step $h$. The system consists
of parallel transversal layers of atoms. In each layer, the
nanotube has $2m$ atoms, which form $m$ elementary cells separated
by the angular distance $\Delta\phi = 2\pi/m$, so that $h$ defines
alternating longitudinal distances between the transverse layers.
Each atom of the CNT can be characterized by three indices
$(n,l,k)$, where $(n,l)$ defines an elementary cell ($n= 0, \pm 1,
\ldots$, $l =1,2, \ldots, m$), and $k$ is the atom number in the
cell, $k=0,1$ (see Fig.~\ref{fig3}).

The Hamiltonian of the lattice of carbon atoms shown in Fig.~\ref{fig3} can
be written in the following general form,
\begin{equation}
{\cal H}=\sum_n\sum_{l=1}^m\biggr[\frac12M(\dot{\bf u}^2_{n,l,0}+\dot{\bf
u}^2_{n,l,1}) + {\cal P}_{n,l}\}\biggr] \label{f21}
\end{equation}
where $M$ is the mass of a carbon atom, $M=12\times 1.6603\cdot
10^{-27}$kg, ${\bf u}_{n,l,k}=(x_{n,l,k}(t),y_{n,l,k}(t),z_{n,l,k}(t))$ is the
radius-vector that defines the position of the carbon atom $(n,l,k)$
in the moment $t$, and the term ${\cal P}_{n,l} \equiv P({\bf
u}_{n-1,l,1},{\bf u}_{n-1,l+1,0},{\bf u}_{n,l,0},{\bf u}_{n,l,1},{\bf
u}_{n+1,l-1,1},{\bf u}_{n+1,l,0})$ denotes the total potential
energy given by a sum of three different types of potentials,
\begin{widetext}
\begin{eqnarray}
&&{\cal P}_{n,l} =V({\bf u}_{n,l,0},{\bf u}_{n,l,1})
+V({\bf u}_{n-1,l+1,0},{\bf u}_{n,l,1})
+V({\bf u}_{n,l,1},{\bf u}_{n+1,l,0})+U({\bf u}_{n-1,l,1},{\bf u}_{n,l,0},{\bf u}_{n,l,1})
\nonumber\\
&+&U({\bf u}_{n+1,l-1,1},{\bf u}_{n,l,0},{\bf u}_{n,l,1})
+U({\bf u}_{n-1,l,1},{\bf u}_{n,l,0},{\bf u}_{n+1,l-1,1})
+U({\bf u}_{n,l,0},{\bf u}_{n,l,1},{\bf u}_{n-1,l+1,0})
\nonumber\\
&+&U({\bf u}_{n,l,0},{\bf u}_{n,l,1},{\bf u}_{n+1,l,0})
+U({\bf u}_{n-1,l+1,0},{\bf u}_{n,l,1},{\bf u}_{n+1,l,0})
+W({\bf u}_{n,l,1},{\bf u}_{n,l,0},{\bf u}_{n-1,l,1},{\bf u}_{n+1,l-1,1})
\label{f22}\\
&+&W({\bf u}_{n,l,1},{\bf u}_{n,l,0},{\bf u}_{n+1,l-1,1},{\bf u}_{n-1,l,1})
+W({\bf u}_{n-1,l,1},{\bf u}_{n,l,0},{\bf u}_{n,l,1},{\bf u}_{n+1,l-1,1})
+W({\bf u}_{n,l,0},{\bf u}_{n,l,1},{\bf u}_{n-1,l+1,0},{\bf u}_{n+1,l,0})
\nonumber\\
&+&W({\bf u}_{n,l,0},{\bf u}_{n,l,1},{\bf u}_{n+1,l,0},{\bf u}_{n-1,l+1,0})
+W({\bf u}_{n-1,l+1,0},{\bf u}_{n,l,1},{\bf u}_{n,l,0},{\bf u}_{n+1,l,0}).
\nonumber
\end{eqnarray}
\end{widetext}

The first three terms describe a change of the deformation energy
due to a direct interaction between pairs of atoms with
coordinates ${\bf u}_1$ and ${\bf u}_2$, characterized by the
potential $V({\bf u}_1,{\bf u}_2)$. The next six terms describe the
deformation energy of the angle between the links ${\bf u}_1{\bf
u}_2$ and ${\bf u}_2{\bf u}_3$, taken into account with the potential
$U({\bf u}_1,{\bf u}_2,{\bf u}_3)$. Finally, the next six terms
describe the deformation energy associated with a change of the
effective angle between the planes ${\bf u}_1{\bf u}_2{\bf u}_3$ and
${\bf u}_2{\bf u}_3{\bf u}_4$, characterized by the potential
$W({\bf u}_1,{\bf u}_2,{\bf u}_3,{\bf u}_4)$.

In our numerical simulations, we employ the interaction potentials
frequently used in modeling of the dynamics of polymer macromolecules~\cite{p1,p2,p3},
\begin{equation}
V({\bf u}_1,{\bf u}_2)=D\{\exp[-\alpha(\rho-\rho_0)]-1\}^2,
\label{ln19}
\end{equation}
where $\rho=|{\bf u}_2-{\bf u}_1|$, $D=4.9632$~eV is the energy of
the valent coupling, and $\rho_0=1.418$\AA  ~is the static length of valent bond;
\begin{equation}
U({\bf u}_1,{\bf u}_2,{\bf u}_3)=\epsilon_v(\cos\varphi-\cos\varphi_0)^2,
\label{ln20}
\end{equation}
where
$$
\cos\varphi=({\bf u}_3-{\bf u}_2,{\bf u}_1-{\bf u}_2)(|{\bf
u}_3-{\bf u}_2|\cdot |{\bf u}_2-{\bf u}_1|)^{-1},
$$

and $\cos \phi_0= \cos (2\pi/3) = -1/2$. Finally,
\begin{equation}
W({\bf u}_1,{\bf u}_2,{\bf u}_3,{\bf
u}_4)=\epsilon_t\left[1-({\bf v}_1,{\bf v}_2)(|{\bf v}_1|\cdot |{\bf
v}_2|)^{-1}\right],
\label{ln21}
\end{equation}
where ${\bf v}_1=({\bf u}_2-{\bf u}_1)\times ({\bf u}_3-{\bf u}_2)$
and ${\bf v}_2=({\bf u}_3-{\bf u}_2)\times ({\bf u}_4-{\bf u}_3)$.
The model parameters such as $\alpha=1.7889$ \AA$^{-1}$, $\epsilon_v=1.3143$ eV, and
$\epsilon_t=0.499$ eV can be determined from the phonon frequency spectrum of
a planar lattice of carbon atoms \cite{p16,pai}.

Equilibrium structure of the nanotube with index $(m,m)$ can be characterized by three
parameters : its radius $R$, the angle shift $\varphi$ and the longitudinal step $h$. Equilibrium
positions of the atoms in the tube are given by the coordinates
\begin{equation}
\begin{tabular}{ll}
$x^0_{n,l,0}=h(n-1)$, & $x^0_{n,l,1}=h(n-1)$, \\
$y^0_{n,l,0}=R\cos(\phi_{n,l})$, &  $y^0_{n,l,1}=R\cos(\phi_{n,l}+\varphi)$,\\
$z^0_{n,l,0}=R\sin(\phi_{n,l})$, &  $z^0_{n,l,1}=R\sin(\phi_{n,l}+\varphi)$,
\end{tabular}
\label{f26}
\end{equation}
with cylindrical angles $\phi_{n,l}=[l-1+(n-1)/2]\Delta\phi$ and the
angular distance $\Delta\phi=2\pi/m$. In order to find the
parameters $R$, $\varphi$ and $h$, we need to solve the minimization
problem
\begin{eqnarray*}
P({\bf u}_{n-1,l,1}^0,{\bf u}_{n-1,l+1,0}^0,{\bf u}_{n,l,0}^0,{\bf u}_{n,l,1}^0,{\bf
u}_{n+1,l-1,1},{\bf u}_{n+1,l,0}^0)\\
\rightarrow \min_{R,\varphi, h},
\end{eqnarray*}
where we have introduced the notations ${\bf
u}_{n,l,i}^0=(x_{n,l,i}^0,y_{n,l,i}^0,z_{n,l,i}^0)$ and $i=0,1$. The
resulting value of the energy is then used as the minimum value. For
a nanotube of the (6,6) type, we find a radius $R=4.1782$ \AA \ and
a longitudinal step $h=1.2590$ \AA \ while for a nano\-tube of the
(12,12) type, one obtains $R=8.3230$ \AA \ and $h=1.2560$ \AA.

If one wishes to study small amplitude vibrations, it is more
convenient to switch to local cylindrical coordinates $u_{n,l,k}$, $v_{n,l,k}$,
$w_{n,l,k}$ defined by
\begin{eqnarray}
x_{n,l,k}&=&x_{n,l,k}^0+u_{n,l,k}, \nonumber\\
y_{n,l,k}&=&y_{n,l,k}^0-v_{n,l,k}\sin\phi_{n,l,k}^0+w_{n,l,k}\cos\phi_{n,l,k}^0,~~~~
\label{f27}\\
z_{n,l,k}&=&z_{n,l,k}^0+v_{n,l,k}\cos\phi_{n,l,k}^0+w_{n,l,k}\sin\phi_{n,l,k}^0,\nonumber
\end{eqnarray}
with the angle $\phi_{n,l,0}=[l-1+(n-1)/2]\Delta\phi$ and
$\phi_{n,l,1}=\phi_{n,l,0}+\varphi$. In this coordinate system, the
Hamiltonian of the carbon nanotube takes the form
\begin{widetext}
\begin{equation}
{\cal H}=\sum_n\sum_{l=1}^m\{\frac12M(\dot{\bf x}_{n,l},\dot{\bf x}_{n,l})
+P({\bf x}_{n-1,l};{\bf x}_{n-1,l+1};{\bf x}_{n,l};{\bf x}_{n+1,l-1};{\bf x}_{n+1,l})\},
\label{f28}
\end{equation}
where the six dimensional vector ${\bf
x}_{n,l}=(u_{n,l,0},v_{n,l,0},w_{n,l,0},u_{n,l,1},v_{n,l,1},w_{n,l,1})$
describes in local coordinates the shifting of the atoms located in
the cell $n,l$ from their equilibrium position.

The equations of motion  for the Hamiltonian (\ref{f28}) are given
by
\begin{eqnarray}
-M\ddot{\bf x}_{n,l}&=&-{\bf F}_{n,l}=
P_{{\bf x}_1}({\bf x}_{n,l};{\bf x}_{n,l+1};{\bf x}_{n+1,l};{\bf x}_{n+2,l-1};{\bf x}_{n+2,l})
+P_{{\bf x}_2}({\bf x}_{n,l-1};{\bf x}_{n,l};{\bf x}_{n+1,l-1};{\bf x}_{n+2,l-2};{\bf x}_{n+2,l-1})
\nonumber\\
&+&
P_{{\bf x}_3}({\bf x}_{n-1,l};{\bf x}_{n-1,l+1};{\bf x}_{n,l};{\bf x}_{n+1,l-1};{\bf x}_{n+1,l})
+P_{{\bf x}_4}({\bf x}_{n-2,l+1};{\bf x}_{n-2,l+2};{\bf x}_{n-1,l+1};{\bf x}_{n,l};{\bf x}_{n,l+1})
\label{f29} \\
&+&
P_{{\bf x}_5}({\bf x}_{n-2,l};{\bf x}_{n-2,l+1};{\bf x}_{n-1,l};{\bf x}_{n,l-1};{\bf x}_{n,l}),
\nonumber
\end{eqnarray}
with the function $P_{{\bf x}_i}
=\frac{\partial}{\partial{\bf x}_i}P({\bf x}_1,{\bf x}_2,{\bf x}_3,{\bf x}_4,{\bf x}_5)$,
$i=1,2,...,5$.
In the linear approximation, the same set of equations takes the
form
\begin{eqnarray} -M\ddot{\bf x}_{n,l}&=&B_1{\bf
x}_{n,l}+B_2{\bf x}_{n+1,l}+B_2^*{\bf x}_{n-1,l}
                      +B_3{\bf x}_{n+2,l}+B_3^*{\bf x}_{n-2,l}
                      +B_4{\bf x}_{n,l+1}+B_4^*{\bf x}_{n,l-1}\nonumber \\
                    &+&B_5{\bf x}_{n+1,l-1}+B_5^*{\bf x}_{n-1,l+1}\label{f30}
                      +B_6{\bf x}_{n+2,l-1}+B_6^*{\bf x}_{n-2,l+1}
                      +B_7{\bf x}_{n+2,l-2}+B_7^*{\bf x}_{n-2,l+2},
\end{eqnarray}
where the matrix coefficients are defined as
\begin{eqnarray*}
B_1=P_{{\bf x}_1{\bf x}_1}+P_{{\bf x}_2{\bf x}_2}+P_{{\bf x}_3{\bf x}_3}+P_{{\bf x}_4{\bf x}_4}
+P_{{\bf x}_5{\bf x}_5},
~~~B_2=P_{{\bf x}_1{\bf x}_3}+P_{{\bf x}_3{\bf x}_5},~~~
B_3=P_{{\bf x}_1{\bf x}_5},\\
B_4=P_{{\bf x}_1{\bf x}_2}+P_{{\bf x}_4{\bf x}_5},~~~
B_5=P_{{\bf x}_2{\bf x}_3}+P_{{\bf x}_3{\bf x}_4},~~~
B_6=P_{{\bf x}_1{\bf x}_4}+P_{{\bf x}_2{\bf x}_5},~~~
B_7=P_{{\bf x}_2{\bf x}_4},
\end{eqnarray*}
\end{widetext}
with the matrix of partial derivatives
$$
P_{{\bf x}_i,{\bf x} j}=\frac{\partial^2 P}{\partial{\bf x}_i\partial{\bf x}_j}
({\bf 0},{\bf 0},{\bf 0},{\bf 0},{\bf 0}),~~~i,j=1,2,3,4,5~.
$$
The solution of the linearized equations (\ref{f30}) can be found in
terms of plane waves in the form
\begin{equation}
{\bf x}_{nl}=A{\bf e}\exp(iqn+il\delta\phi-i\omega t), \label{f31}
\end{equation}
where $A$ stands for the amplitude of the wave, $\bf e$ - the unit
vector of the amplitude, $q\in[0,\pi]$ - the dimensionless wave number and
$\delta=2\pi j/m$ ($j=0,1,...,m-1$) -- the dimensionless
orbital moment of the phonon. By injecting the expression
(\ref{f31}) into the linear equation system (\ref{f30}), we obtain
the eigenvalue problem

\begin{figure}[t,b]
\includegraphics[angle=0, width=1\linewidth]{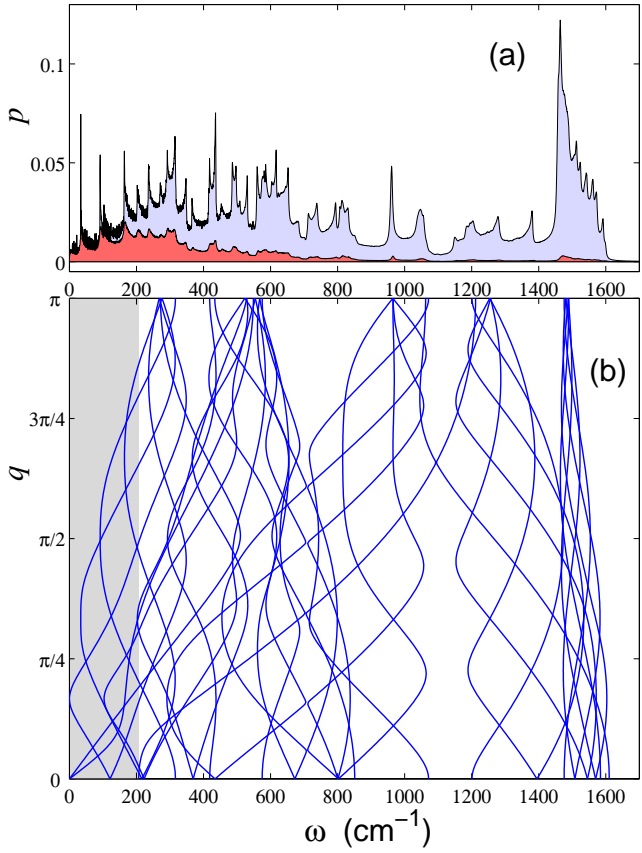}
\caption{
(a) Spectral density of thermal
oscillations of armchair $(6,6)$ carbon nanotube for $T=300$ K
(red field corresponds to system equations with color noise)
and (b) sixty dispersion curves of the phonon modes.
Gray color marks the frequency region $\omega< k_BT/\hbar $.}
\label{fig4}
\end{figure}

\begin{widetext}
\begin{eqnarray}
M\omega^2{\bf A}=[B_1+B_2e^{iq}+B_2^*e^{-iq}+B_3e^{2iq}+B_3^*e^{-2iq}
+B_4e^{i\delta}+B_4^*e^{-i\delta}~~~~\nonumber\\
+B_5e^{iq-i\delta}+B_5^*e^{-iq+i\delta}
+B_6e^{2iq-i\delta}+B_6^*e^{-2iq+i\delta}+B_7e^{2iq-2i\delta}+B_7^*e^{-2iq+2i\delta}]A.
\label{f32}
\end{eqnarray}
\end{widetext}
Thus, the calculation of the dispersion curves of the carbon
nanotube requires the computation of the eigenvalues of the
hermitian matrix of dimension $6\times6$ (\ref{f32}) at each value
of the wave number $0\le q\le \pi$ and the moment $\delta=2\pi j/m$
($j=0,1,...,m-1$). The dispersion curves obtained in this way
consists of $6m$ branches (see Fig. \ref{fig4} (b)).

The computation of the eigenvalues (\ref{f32}) yields not only all the
dispersion curves $\omega(q)$ but also the spectral density
$p(\omega)$, normalized according to $\int_0^\infty
p(\omega)d\omega=1$.

A simple method for obtaining the temperature dependence of the
spectral density consists in making a simulation of the thermal
vibrations of the carbon nanotube. To this aim, the system is first
driven to thermal equilibrium using the usual Langevin equations
with white noise
\begin{equation}
M\ddot{\bf x}_{n,l}={\bf F}_{n,l}-\Gamma M\dot{\bf x}_{n,l}+\Xi_{n,l},
\label{f33}
\end{equation}
with the dissipation coefficient $\Gamma=1/t_r$, $t_r$ - the
relaxation time of atoms (it is reasonable to take $t_r=0.1$ps) and
$\Xi_{n,l}=(\xi_{n,l,1},...,\xi_{n,l,6})$ - the six dimensional
vector corresponding to the normally distributed random noises,
describing the interaction of the particles located in the cell
$(n,l)$ with the thermal bath. The characteristic correlation
functions of these noises can be written as

\begin{equation}
\langle \xi_{n,k,i}(t_1)\xi_{m,l,j}(t_2)\rangle=2Mk_BT\delta_{nm}\delta_{kl}\delta_{ij}
\delta(t_1-t_2), \label{f34}
\end{equation}
where $k_B$ is the Boltzmann constant and $T$ - the temperature of
the thermostat. After having set the initial coordinates (\ref{f26})
and velocities to zero, the numerical integration of the system of
Langevin equations was performed over $t=20\hspace{0.5mm}t_r$. We
then decoupled the thermalized system from the bath and calculated
the spectral density $p(\omega)$ of the kinetic energy distribution
of the atoms by following the real time dynamics of the isolated
system. In order to increase the accuracy of the result, the
spectral density was obtained from 100 independent thermalization
processes and averaged over the atoms of the system.

The spectral density profile obtained at $T=300$ K is shown in Fig.
\ref{fig4} (a). We notice that at this temperature, the density
profile is in good agreement with the shape of the dispersion
curves, which can be seen as a weak manifestation of non-linearity
effects. If one assumes that the proper vibrational modes of the
carbon nanotube remain linear, then the specific heat of the
nanotube can be deduced from the integral

\begin{eqnarray}
c(T)=\int_0^{+\infty}c_q(\omega)p(\omega)d\omega,\label{f35}\\
c_q(\omega)=\left(\frac{\hbar\omega}{k_BT}\right)^2
\frac{\exp(\hbar\omega/k_BT)}{[\exp(\hbar\omega/k_BT)-1]^2} \nonumber
\end{eqnarray}
where $c_q(\omega)$ is the dimensionless thermal capacity of phonons
with angular frequency $\omega$. This method to compute the heat
capacity of carbon nanotubes was first used in Ref. \cite{sh1,sh2}.

\begin{figure}[t,b]
\includegraphics[angle=0, width=1\linewidth]{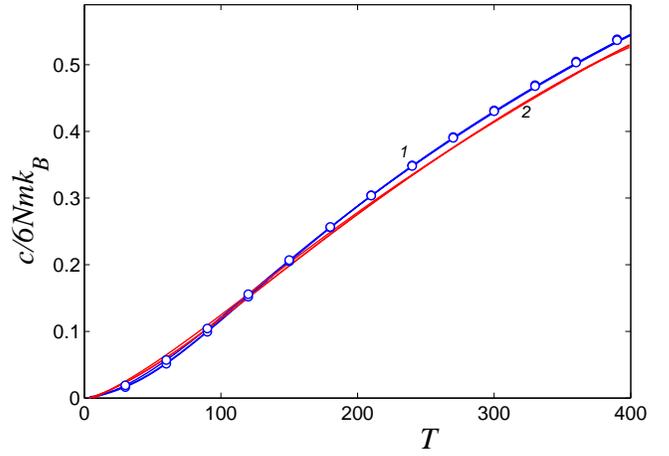}
\caption{
Temperature dependence of specific heat $c$ of carbon nanotube (6,6), (12,12)
and (10,0). Blue line (curve 1) corresponds to the result obtained with the use of the frequency
density of linear phonon waves while markers was obtained from the  frequency density
of thermal vibrations. Red line (curve 2) corresponds to the result obtained
from the numerical integration of
Langevin equations with color noise.}
\label{fig5}
\end{figure}

This approach is remarkably practical since the specific heat of
the system follows from the simple
knowledge of the spectral density $p(\omega)$ of thermal vibrations .
On the other hand, the spectral density can be deduced from the shape of the
spectral curves, that is,  by considering the later as the
frequency spectrum of the harmonic modes in the nanotube.
The heat capacity $c(T)$ of the carbon nanotube with index
(6,6), (12,12) and (10,0) obtained in this way
is shown in Fig. \ref{fig5}. It is clearly seen in this figure that
the specific heat is practically insensitive to the index of the nanotube.
Furthermore, the temperature dependence of the specific heat is linear
in the regime $0<T<400$ K. This temperature behaviour of the specific
heat is in concordance with previous theoretical works
\cite{sh1,sh2} and confirms the experimental measures  of the specific
heat of titanium dioxide nanotubes \cite{sh3}.

We can equally obtain the spectral density that appears in Eq. (\ref{f35})
directly from dynamical simulations, i.e. by following
thermal vibrations of atoms in the carbon nanotube at finite temperature $T>0$.
As it can be checked in Fig. \ref{fig5},
numerical simulations show that these two approaches yield almost the
same result (the spectral density has a weak temperature dependence).

Let us notice that the numerical integration of the Langevin equations with white noise (\ref{f33})
that allows to obtain the dimensionless specific heat of the nanotube $(m,0)$ from the equation
$c(T)=(d\langle H\rangle /dT)/(6Nmk_B)$, where $Nh$ is the length
of the nanotube, shows that the heat
capacity is practically independent of the temperature,
that is, $c=1$ over $0<T<400$ K (this results
from the well-known equipartition theorem of classical statistical mechanics,
which states that the mean energy of each degree
of freedom is equal to $k_BT$). But the situation drastically changes,
if one replaces the white noise of the Langevin equation with a time-correlated
color noise whose temperature dependence is given by the formula
(\ref{f11}). In this case, thermal vibrations of the nanotube
are described by the system of Langevin equations

\begin{eqnarray}
M\ddot{\bf x}_{n,l}&=&{\bf F}_{n,l}-\Gamma M\dot{\bf x}_{n,l}+\Xi_{n,l},
\label{f36}\\
\dot{\Xi}_{n,l} &=& (\Theta_{n,l}-\Xi_{n,l})/t_c,
\label{f37}
\end{eqnarray}
where $\Theta_{n,l}=(\eta_{n,l,1},...,\eta_{n,l,6})$ - the six
dimensional vector corresponding to the normally distributed random
noise and normalized according to (\ref{f34}), the relaxation time
$t_r=1$ ps and the correlation time of random noises,
$t_c=\hbar\sqrt{e-2}/k_BT$.

The numerical integration of the equations of motion (\ref{f36})
and (\ref{f37}) first yields the mean energy
of the nanotube $E=\langle H\rangle$  versus temperature $T$. Then
the specific heat is deduced from the relation
$c(T)=dE/dT$. The result is illustrated in Fig. \ref{fig5}. First of all,
it is clearly seen that the specific heat of nanotubes
(6,6), (12,12) and (10,0) have practically the same temperature behaviour,
that is, it tends to zero for $T\rightarrow 0$ and
it rises  regularly with the increase of the temperature. Furthermore,
the specific heat calculated with color noise coincides very well
with the one obtained from Eq. (\ref{f35}) via computation of the spectral density.

We show in Fig. \ref{fig4}-(a) the shape of the spectral density of
thermal vibrations of the carbon nanotube, obtained with color
noise. We can note that only low frequency vibrations characterized
by $\omega<k_BT/\hbar$ are totally thermalized while the
thermalization of high frequency modes is partial. Moreover, the
degree of thermalization decreases with increasing temperature.

By considering the concrete example of carbon nanotubes, we have
shown that the use of the Langevin equations with color noise
(\ref{f36}) and (\ref{f37}) allows to realize the quantum effect of
partial thermalization of high frequency modes and yields the
correct temperature behaviour of the specific heat for complex
molecular systems. The proposed method becomes very useful
especially for molecular systems with complex configurational
dynamics, for example in the case of macromolecules which possess
globular structure, or simply when one deals with a highly
non-linear dynamics and it becomes meaningless to consider the
existence of a spectral density of small-amplitude linear
vibrations.

Let us mention a further application of the generalized Langevin
approach. The formalism that we have presented consists in coupling
to each particle a color noise of the same amplitude, which drives
the systems to thermal equilibrium. If one instead couples the noise
to chosen particles of the system or applies a noise of different
amplitude to each particle, then the dynamics will be a
non-equilibrium process. In this case, the use of the color noise
may lead to new effects such as ratchet dynamics (see for ex.
\cite{cl1,cl2}), which can be obtained neither within the usual
Langevin approach with white noise nor with a quantum description of
the dynamics.

\section{Conclusions}
We proposed a new method for computing the temperature dependence of
the heat capacity in complex molecular systems. The proposed scheme
is based on the use of the Langevin equation with low frequency
color noise. We showed that the thermal behaviour of the correlation
time of random forces, which is the key characteristic of the
partial thermalization effect, can be described by a linear function
of the inverse bath temperature, that is $t_c=\hbar\sqrt{e-2}/k_BT$.
We next illustrated non-linearity effects by considering two simple
Hamiltonian models and we explicitly showed that the generalized
Langevin approach can be used in the presence of anharmonicities in
the Hamiltonian. Finally, by applying the proposed procedure to
carbon nanotubes, we showed that the consideration of the color
noise in the Langevin equation allows to accurately reproduce the
temperature evolution of the specific heat in many-body systems.

It is well-known that for complex systems having strong
non-linearity effects in the quantum regime ($T<T_E$), there exists
a temperature gap unreachable by existing approximative approaches
such as the self-consistent phonon theory or the spectral density
equations (\ref{f35}), while the drawback of quantum Monte-Carlo
methods is the large numerical cost in the case of many particle
models. The proposed method may be very useful to fill this gap,
especially if one wishes to investigate the thermodynamics of
realistic molecular systems with complex configurational dynamics,
for example in the case of macromolecules which possess globular
structures with a highly non-linear dynamics.

\section*{Acknowledgments}
Alexander Savin thanks the Hong Kong Baptist University for a warm
hospitality during his stay in Hong Kong. This work was supported in
part by the grants of the Hong Kong Research Grants Council and Hong
Kong Baptist University.

\appendix
\section{
The specific heat of the oscillator systems from Kleinert`s
variational Path Integral approach}

We will give in this appendix the heat capacities of the simple
non-linear oscillator systems (\ref{f14}) and (\ref{f16}), obtained
from the first order variational path integral method. Since this
approach has already been intensively discussed \cite{Kleinert,Tognetti}
and applied to several many body problems \cite{TognettiII,Liu,Bao},
we will omit the technical details
of the procedure and only report the analytical form of the effective
potential for each Hamiltonian.

The effective potential obtained by Feynman and Kleinert \cite{Kleinert}
for the quartic quantum oscillator (\ref{f14}) can be written as

\begin{eqnarray}\label{A.1}
W(x)&=&\frac{\beta}{4}x^4+\frac{1}{2}(1+3\beta
a^2)x^2+\frac{3\beta}{4}a^4\nonumber\\
&+&\frac{1-\Omega^2}{2}a^2+T\ln\left[\frac{2T}{\Omega}\sinh\left(\frac{\Omega}{2T}\right)\right],
\end{eqnarray}
where the smearing parameter $a$ and the frequency $\Omega$ are
deduced from the following self-consistent equations :

\begin{eqnarray}\label{A.2}
a^2&=&\frac{1}{2\Omega}\coth\left(\frac{\Omega}{2T}\right)-\frac{T}{\Omega^2},\nonumber\\
\Omega^2&=&1+3\beta(a^2+x^2).
\end{eqnarray}
These equations are solved in ref. \cite{Kleinert} by a numerical
iteration scheme at each $x$ and $T$. Although  this numerical
procedure becomes very complicated when one deals with many body
systems,  it is shown in ref. \cite{Tognetti} that the parameters
$a$, $\Omega$ and the centroid potential $W(x)$ can be obtained for a
one dimensional oscillator chain by Taylor expanding the equations
(\ref{A.2}) with respect to the non-linearity parameter $\beta$ at
the order $0(\beta)$, then substituting the expansions in Eq.
(\ref{A.1}) and keeping only the terms of the same linear order,
which finally yields the centroid potential in a fully analytical
form. In the case of the quartic one-body potential (\ref{f14}),
this procedure yields

\begin{equation}\label{A.3}
W(x)= \frac{\beta}{4}x^4+f_1(T)x^2+f_2(T),
\end{equation}
where we have defined

\begin{eqnarray}\label{A.4}
f_1(T)&=&\frac{1}{2}-\frac{3\beta}{2}
T+\frac{3\beta}{4}\coth\left(\frac{1}{2T}\right)\nonumber\\
f_2(T)&=&\frac{3\beta}{16}\left[2T-\coth\left(\frac{1}{2T}\right)\right]^2\\
&+&T\ln\left[2T\sinh\left(\frac{1}{2T}\right)\right].\nonumber
\end{eqnarray}
The partition function that follows from Eq. (\ref{A.3}) and
(\ref{f15.6}) can now be expressed in terms of the Bessel function
of the second kind as

\begin{equation}\label{A.5}
Z=\sqrt{\frac{f_1}{\beta}}\exp\left(\frac{f_1^2-2\beta f_2}{2\beta
T}\right)\mathbf{K}_{\frac{1}{4}}\left(\frac{f_1^2}{2\beta T}\right)
\end{equation}
and the specific heat is deduced from Eq. (\ref{f15.5}).

The effective potential of the two-body Hamiltonian (\ref{f16})
is computed in a similar way as for one dimensional oscillator chains \cite{Tognetti}.
The trial action that appears in Eq. (\ref{Jensen}) was chosen in the form

\begin{eqnarray}\label{trial}
S_0=\int &d\tau& \left\{\frac{1}{2}\left(\dot{x}^2+\dot{y}^2\right)+\frac{\lambda_x}{2}(x-x_0)^2\right.\nonumber\\
&+&\left.\frac{\lambda_y}{2}(y-y_0)^2+g(x-x_0)(y-y_0)\right\}.
\end{eqnarray}
where $\lambda_x$, $\lambda_y$  and $g$ denote the trial parameters
that minimize the Jensen-Peirels inequality (\ref{Jensen}). After
going through the usual minimization procedure and expanding the
smearing parameter $a$ and the centroid potential at the order
$0(\beta)$, one obtains

\begin{eqnarray}\label{centroid}
a_0^2=&-&T\frac{\omega_x^2+\omega_y^2}{\omega_x^2\hspace{0.5mm}\omega_y^2}\nonumber\\
&+&\frac{1}{2\omega_x\hspace{0.5mm}\omega_y}\left\{\omega_x\coth\left(\frac{\omega_y}{2T}\right)+
\omega_y\coth\left(\frac{\omega_x}{2T}\right)\right\}\nonumber,\\
W(x,y)=&-&T\ln\left\{\frac{\omega_x\omega_y}{4T^2}\frac{1}{\sinh(\omega_x/2T)\sinh(\omega_y/2T)}\right\}\nonumber\\
&+&\frac{\omega_x^2}{2}x^2+\frac{\omega_y^2}{2}y^2+\frac{3\beta}{4}a_0^4\nonumber\\
&+&\frac{\beta}{4}(x-y)^4+\frac{3\beta}{2}a_0^2(x-y)^2.
\end{eqnarray}
The integration in Eq. (\ref{f15.6}) can be carried out analytically by performing the
coordinate transformation

\begin{equation}\label{transfo}
u=\frac{x-y}{\sqrt2},\hspace{1cm}v=\frac{x+y}{\sqrt2}.
\end{equation}
We finally obtain for the partition function

\begin{eqnarray}\label{partII}
Z&=&\frac{\Omega\hspace{0.5mm}\omega_x\omega_y}{8\sqrt{\pi\beta T(\omega_x^2+\omega_y^2)}}\frac{1}{\sinh(\omega_x/2T)\sinh(\omega_y/2T)}\nonumber\\
&\times&\exp\left(-\frac{6\beta^2a_0^4+\Omega^4}{8\beta T}\right)\mathbf{K}_{\frac{1}{4}}\left(\frac{\Omega^4}{8\beta T}\right)
\end{eqnarray}
where we have defined

\begin{equation}\label{omega}
\Omega^2=\frac{\omega_x^2\hspace{0.5mm}\omega_y^2}{\omega_x^2+\omega_y^2}+3\beta a_0^2.
\end{equation}

\end{document}